\documentclass[a4paper, 12pt]{article}

\usepackage{amsmath,amsthm,amssymb,amsfonts,amscd,latexsym}
\usepackage{graphicx,subfigure}
\usepackage{tabularx}
\newcommand{\Cov} {\mbox{$\rm{Cov}$\,}}
\newcommand{\Var} {\mbox{$\rm{Var}$\,}}
\newcommand{\Prob} {\mbox{$\rm{Pr}$\,}}

\title{Characterising the $D_2$ statistic: word matches in biological sequences}

\begin{document}

\maketitle

\begin{center}
\textbf{Sylvain For\^et, Susan R.\ Wilson, Conrad J.\ Burden}
\end{center}

\small{
Mathematical Sciences Institute, The Australian National University, Canberra ACT 0200,
Australia\\
}

\begin{abstract}
Word matches are often used in sequence comparison methods, either as a measure
of sequence similarity or in the first search steps of algorithms such as BLAST
or BLAT.  The D2 statistic is the number of matches of words of k letters
between two sequences.  Recent advances have been made in the characterisation
of this statistic and in the approximation of its distribution.  Here, these
results are extended to the case of approximate word matches.

We compute the exact value of the variance of the D2 statistic for the case of
a uniform letter distribution, and introduce a method to provide accurate
approximations of the variance in the remaining cases.  This enables the
distribution of D2 to be approximated for typical situations arising in
biological research.  We apply these results to the identification of
cis-regulatory modules, and show that this method detects such sequences with a
high accuracy.

The ability to approximate the distribution of D2 for both exact and
approximate word matches will enable the use of this statistic in a more
precise manner for sequence comparison, database searches, and identification
of transcription factor binding sites.
\end{abstract}

\section{Introduction}

Alignment-free sequence comparison methods based on word matches allow
sequences to be compared without assuming contiguity of homologous segments.
This is of particular interest for the comparison of biological sequences,
where deletions, insertions or duplications of segments are common.
Several such methods have thus been implemented (see \cite{HHL+08}, for
example), and have had various applications, such as the clustering of large
EST databases (for example, \cite{CvG+01}.
These applications, however, typically rely on empirical thresholds, rather
than on rigorous statistical theory.

One of the statistics for alignment free sequence comparison that has received
much attention is the $D_2$ statistic, which measures the number of words
shared between two sequences.
The characterisation of this statistic started with the calculation of its
mean, and with approximations to the variance \cite{Wat95}.
Later, more accurate approximations of the variance allowed
asymptotic regimes of $D_2$ to be derived for non-uniform \cite{LHW02} and
uniform \cite{KBB+06} letter distributions.
More recently, the exact value of the $D_2$ variance has been
computed \cite{KRS07,FSWSR09}.
In parallel with this theoretical effort, optimal word sizes for typical
biological situations were computed \cite{FKB06}, and practical approximations
of the distribution of $D_2$ in these settings were proposed \cite{FSWSR09}.

A more general version of the $D_2$ statistic is the number of approximate word
matches between two sequences.
After an initial characterisation of the mean of this statistic, an asymptotic
distribution regime was characterised when the logarithm of the sequence size
is large compared with the word size \cite{BKW08}.
Here, we further characterise the $D_2$ statistic in the case of approximate
word matches, by computing its variance and proposing approximations of its
distribution for typical biologically relevant situations.
Finally, we present an application of these results to the identification of
regulatory sequences.

\section{Results}

\subsection{Definitions}

The statistic $D_2(n_A, n_B, k, t, \eta)$ ($D_2$ henceforth) is the number of
approximate word matches of length $k$ with up to $t$ mismatches between
sequences
$A = (A_1 \ldots A_{n_A})$
and
$B = (B_1 \ldots B_{n_B})$
with $A_i$ and $B_j$ belonging to an alphabet $\mathcal{A}$ and distributed
according to a letter distribution parameterised by $\eta$.
As previously \cite{FSWSR09}, for mathematical convenience we will impose
periodic boundary conditions, that is, the letter in the first position in a
sequence is assumed to follow the last letter of that sequence.
Also, only the case of strand symmetric Bernoulli text will be considered, that
is, sequences built from alphabets of four iid (independent and identically
distributed) letters (A, T, G and C) with the further constraint that the
probability $\xi_a$ of letter $a \in \mathcal{A}$ occurring is
$\xi_A = \xi_T = \frac{1}{4}(1 + \eta)$
and
$\xi_G = \xi_C = \frac{1}{4}(1 - \eta)$, where $0 \le \eta \le 1$.
Note that the periodic boundary conditions simplify the theoretical
calculations considerably, but allow the method to be used for linear
as well as circular sequences by appropriate preprocessing (see
Section 2.5 for example).

Defining the $t$ neighbourhood match indicator
\begin{equation}
    Y_{(i,j)} = \left\{
    \begin{array}{ll}
        1 & \mbox{if } \Delta\left((A_i, \ldots, A_{i + k - 1}), (B_j, \ldots, B_{j + k - 1})\right) \leq t \\
        0 & \mbox{otherwise}
    \end{array}
    \right.
\end{equation}
where $\Delta(w_1, w_2)$ is the number of mismatches between the words $w_1$
and $w_2$, the $D_2$ statistic is given by
\begin{equation}
    D_2 = \sum_{(i,j) \in I} Y_{(i,j)}
    \label{eq::d2Definition}
\end{equation}
where the index set is $I = \{(i, j) : 1 \le i \le n_A, 1 \le j \le n_B\}$.

\subsection{$D_2$ mean}

The mean of $D_2$ was first computed for exact word matches ($t = 0$) and iid
letters \cite{Wat95}.
This was later extended to the case of letters generated by a Markov
model \cite{KRS07}.
A formula for the mean was also computed for approximate word matches
($t \geq 0$) in the case of Bernoulli symmetric text \cite{BKW08} in terms
of the perturbed binomial distribution \cite{MM04}.
In Appendix~\ref{section:mean} we derive the equivalent formula
\begin{equation}
E\left[D_2\right] = \frac{n_A n_B}{4^k} \sum_{l = 0}^t {k \choose l} (3 - \eta^2)^l (1 + \eta^2)^{k - l}.
\end{equation}

\subsection{$D_2$ variance}

An exact formula for the variance of $D_2$ was derived in the case of iid
letters and exact word matches using periodic boundary conditions in
\cite{FSWSR09}.
Another study computed the variance for exact word matches using free boundary
conditions, in the cases of iid letters and of letters generated by a Markov
model \cite{KRS07}.
Here we extend these results to the case of approximate word matches for iid
letters and Bernoulli symmetric text, using periodic boundary conditions.
Specific details of this technical derivation are given in
Appendix~\ref{section:var}.
A brief summary is given below.

To calculate the variance of $D_2$ for approximate word matches and symmetric
Bernoulli text, we follow the method used in \cite{FSWSR09}.
First we deduce from equation~(\ref{eq::d2Definition}) that:
\begin{equation}
    \Var(D_2) = \Var \left( \sum_{(i,j) \in I} Y_{(ij)} \right)
              =  \sum_{(i,j)\in I} \sum_{(i',j') \in I} \Cov(Y_{(ij)}, Y_{(i'j')}).
\end{equation}
We set $u = (i, j)$, $v = (i', j')$ and for fixed $u$ split the sum over $v$ as follows.
Let
$J_u = \{v = (i',j') : |i'-i| < k \mbox{ or } |j'-j| < k \}$
be the dependency neighbourhood of $Y_u$.  For $v \notin J_u$, $\Cov(Y_u, Y_v) = 0$.
$J_u$ is decomposed into two disjoint sets \cite{Wat95}:
an accordion set,
$J_u^a = \{v = (i',j') : |i'-i| < k \mbox{ and } |j'-j| < k \}$
(when two pairs of matching words overlap in both sequences)
and a crabgrass set,
$J_u^c = J_u \setminus J_u^a$
(when two pairs of matching words overlap in one sequence only).
The accordion set is further decomposed into
a diagonal part,
$J_u^{ad} = \{v = (i',j') : -k < i'-i = j'-j < k\}$
and an off-diagonal part,
$J_u^{ao} = J_u^a \setminus J_u^{ad}$.

Table~\ref{table::variance} gives a summary of the components of the variance
in different settings.
The only case that is not analytically characterised is the off-diagonal part
of the accordion for approximate word matches and non-uniform letter
distribution.
In this case, however, numerical tables can be assembled to approximate the
entire accordion part of the variance with good accuracy.
To see this, note that the accordion part takes the form
$n_An_B\Phi(k, t, \eta)$.
When $n_A = n_B = 2k - 1$, the only index set contributing to the variance
is the accordion part.
Although computing $D_2$ for approximate word matches requires an algorithm
with complexity $o(n_A n_B)$, it is relatively inexpensive to approximate the
variance of $D_2$ by simulation for small $n_A$ and $n_B$.
Tables of the function $\Phi$ were thus approximated by simulating a large
number of pairs of sequences of length $2k - 1$ for $k \leq 16$ and setting
$\Phi(k, t, \eta) = \Var(D_2(2k - 1, 2k - 1, k, t, \eta))/(2k - 1)^2$(see
below).

\begin{table}

    \newcolumntype{Y}{>{\setlength{\hsize}{1.0\hsize}\centering\arraybackslash}X}
    \newcolumntype{Z}{>{\setlength{\hsize}{0.5\hsize}\centering\arraybackslash}X}
    \renewcommand{\tabularxcolumn}[1]{m{#1}}

    \begin{tabularx}{\linewidth}{Y Z Z Z Z}
                                        & crabgrass & accordion, diagonal           & accordion, off-diagonal \\
    \hline
    exact matches, uniform distribution
    ($t = 0$, $\eta = 0$)               & 0         & Eq.~(20) of \cite{KBB+06}     & 0                       \\
    \hline
    exact matches, non-uniform distribution
    ($t = 0$, $\eta \neq 0$)            & Eq.~(14) of \cite{FSWSR09} & Eq.~(17) of \cite{FSWSR09}  & Eqs.~(20) and (26) of \cite{FSWSR09} \\
    \hline
    approximate matches, uniform distribution
    ($t \neq 0$, $\eta = 0$)            & 0         & Appendix~\ref{section:unifacc} & 0                       \\
    \hline
    approximate matches, non-uniform distribution
    ($t \neq 0$, $\eta \neq 0$)         & Appendix~\ref{section:nonunifcrab}         & \multicolumn{2}{c}{Appendix~\ref{section:nonunifacc}} \\
    \hline
    \end{tabularx}

    \caption{Contribution of the index sets of the dependency neighbourhood to
             the variance of $D_2$. See text for definitions.}
    \label{table::variance}
\end{table}

\subsection{$D_2$ distribution}

It has been shown previously \cite{FSWSR09} that for exact word matches and in
most biologically relevant situations, a distribution chosen \emph{ad-hoc}
such as the gamma distribution can provide a better estimate of the $D_2$
distribution than the asymptotic normal distribution.
Here we provide approximations for the distribution of $D_2$ in the case of
approximate word matches.

For convenience we have set $n_A = n_B = n$ in our numerical simulations.
We have simulated the distribution of $D_2$ for sequence sizes ranging from small
ESTs ($n =100$) to reasonably large genes ($n = 3200$), for even word
sizes $k$ between 2 and 16, for every possible number of mismatches
($0 \leq t < k$), and for both uniform ($\eta = 0$) and non-uniform
($\eta = \frac{1}{3}$) letter distributions.
For each combination of parameters, $10^6$ pairs of iid sequences were
generated.
Tables of the accordion contribution function $\Phi$ were
estimated by generating $10^9$ pairs of iid sequences of size $n = 2k - 1$, with $k$ ranging
from 2 to 16 with an increment of 2.
The Mersenne-Twister random number generator \cite{MMNT98} was used, as
implemented in the GNU scientific library ({\tt http://www.gnu.org/software/gsl/}).
The code was written in ANSI C and is available from the authors'
website ({\tt http://wwwmaths.anu.edu.au/cbis/\~{}sf/k\_words}).

Previously, the gamma distribution was used to approximate the $D_2$
distribution in the case of exact word matches \cite{FSWSR09}.
Here, the beta distribution scaled to the range $[0, n^2]$ is used instead of
the gamma distribution.
In the range of parameters assessed in our simulations, the gamma and beta
distributions are mostly indistinguishable (data not shown).
It might be expected, however, that the beta distribution provides better
approximations for very small p-values, as it is bounded within the same domain
of definition as $D_2$ ($[0, n_An_B]$), whereas the gamma distribution is
defined from zero to infinity.
Histograms of our numerical simulations the $D_2$ statistic are compared with
the density function of the beta distribution scaled to this interval, that is
\begin{equation}
\frac{1}{n_A n_B}f_B\left(\frac{x}{n_A n_B}; \alpha, \beta\right)
\end{equation}
where
$f_B(x; \alpha, \beta) = \Gamma(\alpha + \beta)/(\Gamma(\alpha)\Gamma(\beta))x^{\alpha - 1}(1 - x)^{1 - \beta}$
is the canonical density function of the beta distribution.
The parameters $\alpha$ and $\beta$ are set so that the mean and variance of
the scaled beta distribution agree with the theoretical values
$\mu = E[D_2]$, $\sigma^2 = \Var(D_2)$ derived in the appendix:
\begin{equation}
\alpha = \frac{\mu}{n_A n_B}\left[ \frac{\mu(n_A n_B - \mu)}{\sigma^2} - 1 \right],
\qquad
\beta = \frac{n_A n_B - \mu}{n_A n_B}\left[ \frac{\mu(n_A n_B - \mu)}{\sigma^2} - 1 \right]. \\
\end{equation}

\begin{figure}
    \includegraphics[width=\textwidth]{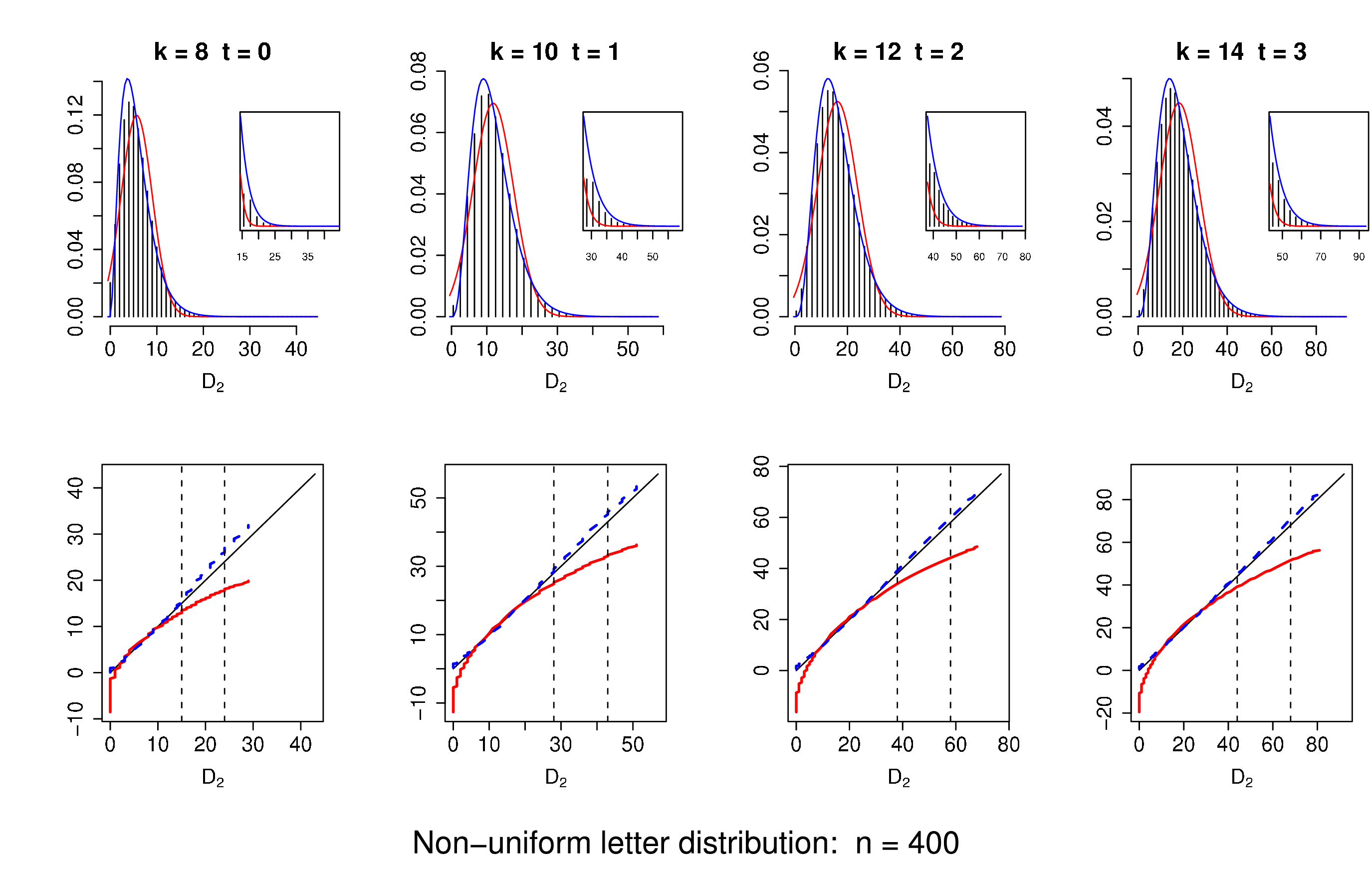}
    \caption{\emph{Top row}: Histograms of the simulated distribution of $D_2$
             for sequences of size $n = 400$ and non-uniformly distributed
             letters. The normal distribution is shown in red and the beta
             distribution in blue.
             The insert shows a close up on the far right of the tail
             larger than the $99^{th}$ percentile.
             \emph{Bottom row}: quantile-quantile plots with the simulated
             $D_2$ values horizontally, and the normal (continuous red
             line) and beta (dashed blue line) values vertically. The
             vertical dashed lines represent the 0.99 and the 0.9999
             quantiles.}
    \label{figure::distAndQQPlots}
\end{figure}

Figure~\ref{figure::distAndQQPlots} shows the simulated distribution of $D_2$
for the size typical of a small EST or a read produced by the 454 Titanium
technology (sequence size $n = 400$), in the case of non-uniform letter
distributions ($\eta = \frac{1}{3}$).
The word sizes displayed in this figure are the optimal word sizes
corresponding to the associated number of mismatches.
We use the optimal word sizes computed previously in \cite{FKB06}.
In brief, a word size and number mismatches combination is optimal when it best
captures the relatedness between artificially evolved sequences using the $D_2$
statistic as a relatedness estimator.

The quantile-quantile plots between the beta and normal distributions, and the
simulated $D_2$ distribution show unambiguously that for these parameters
combinations, the beta distribution provides a closer fit to the $D_2$
distribution than the normal distribution.
Similar figures for all the simulations can be found on the authors' website
({\tt http://wwwmaths.anu.edu.au/cbis/\~{}sf/k\_words}).
We observed a few rare situations where the normal distribution outperformed
the beta distribution, but these were cases where the number of mismatches was
close to the word size, and are of little practical importance.

\subsection{Application to the detection of regulatory sequences}

We now apply the approximation of the $D_2$ distribution to a practical
biological problem: the identification of sequences containing
\emph{cis-regulatory modules} (CRMs).

We use the same dataset as \cite{KRS07}, which contains seven sets of
sequences known to contain CRMs.
Within each set, the CRMs are driving gene expression in one particular tissue
or life stage.
The sets contain between 9 and 82 sequences.
For each of these `positive' sets, a `negative' set was constructed from
randomly chosen non-coding sequences of the same species, containing the same
number of sequences and with the same sequence sizes as in the positive set.

In \cite{KRS07}, the authors primarily assessed whether their method can
capture an expected effect, namely that sequences known to contain similar
(CRMs) are more related to each other than are randomly selected sequences.
While they show that the $D_2$ based approach clearly outperforms other
techniques, this approach is of limited practical use.

We chose instead to address a problem more frequently faced by practitioners:
given a set of sequences known to contain CRMs, and a query sequence, can the
query sequence be classified as containing similar CRMs or not?
We set up the following experiment: each sequence in each positive set was
selected as the query sequence and compared both to the remaining positive
sequences of this set and to the corresponding negative sequences.
In order that our theoretical results for the iid hypothesis null distribution
could be applied, each sequence was preprocessed by (1) joining the ends to
effect periodic boundary conditions and (2) removing masked tandem repeats
present in the data sets and concatenating the pieces either side of the
removed portion.
The parameters $n_A$ and $n_B$ were taken from the preprocessed sequences and
for each pairwise comparison the parameter $\eta$ estimated from the combined
letter frequencies of the two sequences in question.
The query sequences were then screened to accept only those for which the
smallest smallest p-value of all comparisons was less than 0.01.
We used a stringent criterion, namely, a positive query sequence was considered
correctly classified if the smallest p-value was obtained with another sequence
of the positive set.

\begin{figure}
    \includegraphics[width=\textwidth]{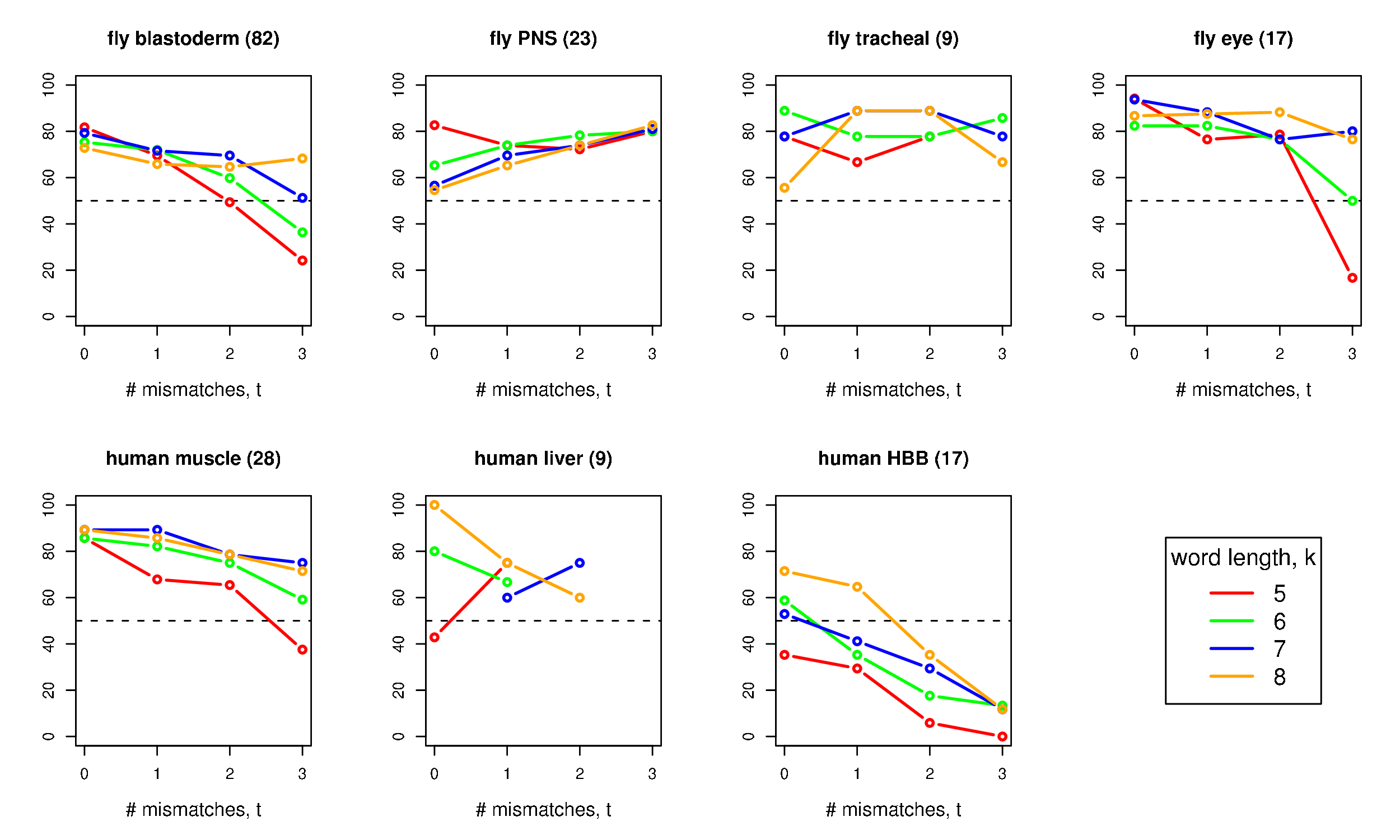}
    \caption{Percentage of times that a sequence containing CRMs is correctly
             classified: each subplot corresponds to a type of CRM, and the
             numbers in parentheses are the number of positive control
             sequences in each set.  Percentages are only plotted if at least 4
             query sequences survived the screening requirement that the
             minimum p-value should be less than 0.01.}
    \label{figure::classificationPos}
\end{figure}

Figure \ref{figure::classificationPos} shows the results of this experiment.
A good sensitivity is achieved in most datasets, with typically 80\% or more of
the sequences correctly classified for at least one parameter combination using
this stringent criterion.
The optimal parameters vary from one condition to another.
This may reflect different properties of the underlying CRMs, in terms of size,
letter composition and level of conservation that they require in order to be
functional.
The problem of choosing optimal parameters is easily solved by using the above
approach, namely by determining a set of positive sequences and using these to
estimate appropriate parameters before comparing the query sequence(s) to them.

The percentage of correctly classified negative sequences based on the
smallest p-value was typically around 50\% (data not shown).  This
suggests that while this method can successfully identify candidates,
further validation of the candidates would be needed.

\section{Discussion}

In this study we present exact values and approximations of the variance of
$D_2$ for pairs of symmetric Bernoulli texts.
These results enable the distribution of $D_2$ to be approximated with or
without mismatches for most situations occurring in biological research.

We illustrate the application of these results by using the $D_2$ statistic to
identify sequences containing regulatory modules.  Our results show that this
method can be used to identify candidate regulatory sequences for further
experimental validation, or in combination with other prediction methods.

A remaining theoretical problem is evaluation of the variance and distribution
of the $D_2$ statistic in the case of approximate word matches for strings that
are not symmetric Bernoulli texts, such as proteins.
This lack of theory could be partially circumvented by using exact word matches
for protein searches, but using alphabet reduction to account for most common
substitutions.
A similar alphabet reduction resulted in increased accuracy in the construction
of phylogenetic trees with an alignment free method \cite{HR07}.

\section*{Appendix}

\appendix

\section{Derivation of D2 mean and variance}

Define the statistic $D_2$ to be the number of $k$-word matches with up to $t$
mismatches ($t = 0, \ldots, k$) between sequences A and B of letters drawn from
an alphabet $\mathcal{A}$.
Let the sequence lengths be $n_A$ and $n_B$ respectively, and assume each
sequence to consist of i.i.d.\ random letters with probability $\xi_a$ of
letter $a \in \mathcal{A}$ occurring at any given location, where
$\sum_{a \in \mathcal{A}}\xi_a = 1$.
Also assume periodic boundary conditions on both sequences, that is, the letter
in the first position in sequence A is assumed to follow the letter in the
${n_A}^{\rm th}$ position, and the letter in the first position in sequence B
is assumed to follow the letter in the ${n_B}^{\rm th}$ position.

In general, we restrict ourselves to the case of strand symmetric Bernoulli
texts of nucleotide sequences, that is, i.i.d.\ sequences for which
$\xi_C = \xi_G = \frac{1}{4}(1 - \eta)$, $\xi_A = \xi_T = \frac{1}{4}(1 + \eta)$,
where $0 \le \eta \le 1$, and write the $D_2$ statistic as
$D_2(n_A, n_B,k,t,\eta)$.

\subsection{Preliminaries}

We use the following terminology adapted from \cite{BKW08}:
\begin{enumerate}
\item For $m = 1, 2, \ldots$, define
$p_m = \sum_{a \in \mathcal{A}} {\xi_a}^m$.
For strand symmetric Bernoulli texts, $p_ 2 = (1 + \eta^2)/4$.
\item Define $\Delta(\mathbf{W}_1, \mathbf{W}_2)$ to be a random variable equal
to the number of mismatches between the two random $k$-words $\mathbf{W}_1$ and
$\mathbf{W}_2$.
When there is no possibility of confusion, we simply write $\Delta(k)$ for the
number of mismatches between the two random $k$-words.
One easily checks that $\Delta(k)$ is a binomial random variable:
\begin{eqnarray}
\Prob(\Delta(k) = l) & = & \Prob(\mbox{Exactly $l$ mismatches and $k - l$ matches}) \nonumber \\
      & = & {k \choose l} (\mbox{prob. of mismatch})^l (\mbox{prob. of match})^{k -l}       \nonumber \\
      & = & {k \choose l} (1 - p_2)^l {p_2}^{k - l}      \nonumber \\
      & = & {k \choose l} \frac{1}{4^k} (3 - \eta^2)^l (1 + \eta^2)^{k - l}.
\label{DeltaDistribution}
\end{eqnarray}
\item $Y_{(i,j)} = Y_u = $ the approximate word match indicator, taking the
value 1 if the number of mismatches between $k$-word at $i$ in $A$ and the
$k$-word at $j$ in $B$ is at most $t$.
That is:
\begin{equation}
    Y_{(i,j)} = \left\{
    \begin{array}{ll}
        1 & \mbox{if } \Delta((A_1, \ldots, A_{i + k - 1}), (B_1, \ldots, B_{j + k - 1})) \leq t \\
        0 & \mbox{otherwise}
    \end{array}
    \right.
    \label{eq::indicatorVariable}
\end{equation}
Note that $ D_2 = \sum_{i = 1}^{n_A} \sum_{j = 1}^{n_B} Y_{(i,j)}$.
\item $g_t(k, \eta, c)$, $G_t(k, \eta, c)$, probability and cumulative
distribution functions of the perturbed binomial distribution \cite{MM04}.
Given a fixed $k$-word with CG-content $c$ ($c = 0, \ldots, k$),
$g_t(k, \eta, c)$ (resp.\ $G_t(k, \eta, c)$)
is the probability that the number of mismatches between that word and a random
$k$-word will be equal to (resp.\ at most) $t$.
Specifically:
\begin{eqnarray}
G_t(k, \eta, c) & = & \sum_{r = 0}^t g_r(k, \eta, c) \\
g_t(k, \eta, c) & = & h(k, \eta, c) u_t(k, \eta, c),
\end{eqnarray}
where $0 \le c$, $t \le k$ are integers, and
\begin{eqnarray}
h(k, \eta, c) & = & \frac{1}{4^k}(1 - \eta)^c(1 + \eta)^{k-c} \\
u_t(k, \eta, c) & = & \sum_{i=0}^{k-t} {c \choose i} {{k - c} \choose {k - t - i}} v_t(i, \eta, c) \\
v_t(i, \eta, c) & = & \left(\frac{3 + \eta}{1 - \eta}\right)^{c - i} \left(\frac{3 - \eta}{1 + \eta}\right)^{t - c + i}.
\end{eqnarray}
In the above definition, we follow a convention that
${n \choose a} = 0$ if $a < 0$ or $a > n$.
\item We set $I = \{(i, j) \, : \,  1 \le i \le n_A, 1 \le j \le n_B \}$.
Given $u = (i, j) \in I$, the dependency neighbourhood of $u$ is defined as:
\begin{equation}
J_u=\{v=(i',j') \, : \,  |i'-i| < k \, \mbox{ or }\, |j'-j| < k \}.
\end{equation}
Note that for $v \notin J_u$, $\Cov(Y_u, Y_v) = 0$.
$J_u$ is divided into two parts, accordion $J_u^a$ and crabgrass $J_u^c$
defined by
\begin{eqnarray}
J_u^a & = & \{v=(i',j') \in J_u \, : \,  |i'-i| < k \, \mbox{ and }\, |j'-j| < k \} \nonumber \\
J_u^c & = & J_u \setminus J_u^a.
\end{eqnarray}
The accordion set is further decomposed into a diagonal part, $J_u^{ad}$ and an
off-diagonal part, $J_u^{ao}$:
\begin{eqnarray}
J_u^{ad} & = & \{v = (i',j') : -k < i'-i = j'-j < k \} \\
J_u^{ao} & = & J_u^a \setminus J_u^{ad}.
\end{eqnarray}
\end{enumerate}

\subsection{Mean of $D_2$}
\label{section:mean}

An equivalent and more concise formula for $E[D_2]$ to that given in
\cite{BKW08} is
\begin{eqnarray}
E[D_2(n_A, n_B, k, t, \eta)] & = & \sum_{(i, j) \in I}  E[ Y_{(i,j)} ]          \nonumber \\
                     & = & n_A n_B \sum_{l = 0}^t \Prob(\Delta(k) = l)   \nonumber \\
                     & = & n_A n_B \sum_{l = 0}^t {k \choose l} (1 - p_2)^l {p_2}^{k - l}  \nonumber \\
                     & = &  \frac{n_A n_B}{4^k} \sum_{l = 0}^t {k \choose l} (3 - \eta^2)^l (1 + \eta^2)^{k - l}.
\end{eqnarray}

\subsection{Variance of $D_2$}
\label{section:var}

An exact formula for the variance of $D_2(n_A, n_B, k, 0, \eta)$ (i.e.\ the
case of exact word matches) has previously been given by \cite{FSWSR09}.
The case of approximate word matches, $0 \le t \le k$, is dealt with here.
We have
\begin{eqnarray}
\Var(D_2(n_A, n_B, k, t, \eta)) & = & \Var \left( \sum_{u \in I} Y_u \right) \nonumber \\
            &  =  & \sum_{u \in I} \sum_{v \in J_u^c} \Cov(Y_u, Y_v)
                      + \sum_{u \in I} \sum_{v \in J_u^a} \Cov(Y_u, Y_v) \nonumber \\
             & = & \left. \Var (D_2) \right|_{\rm crabgrass} + \left. \Var (D_2) \right|_{\rm accordion}.
    \label{eq::varD2}
\end{eqnarray}
Below we give an exact formula for the crabgrass part.
A convenient exact formula for the accordion part remains intractable in
general, and we give below a practical alternate numerical method for its
evaluation.
For the case of a uniform letter distribution, $\eta = 0$, we demonstrate below
(in section~\ref{section:unif}) that only the diagonal part of the accordion
contributes to the variance of $D_2$, and give an exact formula for this case.

\subsubsection{Crabgrass contribution to $\Var (D_2)$}
\label{section:nonunifcrab}

From Eqs.~(6) and (7) on page 9 of \cite{BKW08}, the crabgrass contribution is given by
\begin{eqnarray}
\left. \Var (D_2) \right|_{\rm crabgrass} & = &
                                   \sum_{u \in I} \sum_{v \in J_u^c} \Cov(Y_u, Y_v)  \nonumber \\
                             & = & n_A n_B (n_A + n_B - 4k + 2) \sum_{r = -k + 1}^{k - 1} \Var(f_{|r|}({\mathbf W})), \nonumber \\
\label{crabVar}
\end{eqnarray}
where, for a given $(k - r)$-word ${\mathbf w} \in \mathcal{A}^{k - r}$,
\begin{eqnarray}
f_r({\mathbf w}) & = & \sum_{l = 0}^{\min (r,t)} \Prob(\Delta(r) = l) G_{t - l}(k - r, \eta, c_{\mathbf w})
                                                                          \nonumber \\
          & = & \sum_{l = 0}^{\min (r,t)} {r \choose l} (1 - p_2)^l {p_2}^{r - l} G_{t - l}(k - r, \eta, c_{\mathbf w})
                                                                          \nonumber \\
          & = & \sum_{l = 0}^{\min (r,t)} {r \choose l} \frac{(3 - \eta^2)^l (1 + \eta^2)^{r - l}}{4^r} G_{t - l}(k - r, \eta, c_{\mathbf w}),
\label{frDef}
\end{eqnarray}
where $c_{\mathbf w}$ is the GC-content of $\mathbf w$.
The variance with respect to the random $(k - r)$-word ${\mathbf W}$ is
calculated using
\begin{equation}
\Var(f_r({\mathbf W})) = E[f_r({\mathbf W})^2] - E[f_r({\mathbf W})]^2.
\end{equation}
Since the ${\mathbf w}$-dependence of the function $f_r$ is only via the
GC-content of ${\mathbf w}$, the expectation values are calculated using
\begin{eqnarray}
E[\phi(c_{\mathbf W})] & = & \sum_{c = 0}^{k - r} \Prob(c_{\mathbf W} = c) \phi(c) \nonumber \\
     & = & \sum_{c = 0}^{k - r}  {{k - r}\choose c} (\xi_C + \xi_G)^c (\xi_A + \xi_T)^{k - r- c} \phi(c) \nonumber \\
     & = & \sum_{c = 0}^{k - r}  {{k - r}\choose c} \frac{1}{2^{k - r}} (1 - \eta)^c (1 + \eta)^{k - r - c} \phi(c).
\end{eqnarray}

\subsubsection{Accordion contribution to $\Var (D_2)$}
\label{section:nonunifacc}

The accordion part is
\begin{eqnarray}
\left. \Var (D_2) \right|_{\rm accordion}  & = &
                                   \sum_{u \in I} \sum_{v \in J_u^a} \Cov(Y_u, Y_v),  \nonumber \\
          & = & n_A n_B \Phi(k,  t, \eta),
\end{eqnarray}
where
\begin{equation}
\Phi(k,  t, \eta) = \sum_{r = -k + 1}^{k - 1} \sum_{s = -k + 1}^{k - 1} \Cov(Y_u, Y_{u + (r,s)})
\end{equation}
is independent of $n_A$ and $n_B$.
For the case $n_A = n_B = 2k - 1$, Eq.~(\ref{crabVar}) implies
$\Var (D_2) |_{\rm crabgrass} = 0$, giving
\begin{equation}
\Phi(k, t, \eta) = \frac{\Var \left(D_2(2k - 1, 2k - 1, k, t, \eta) \right)}{(2k - 1)^2},
\end{equation}
which can be estimated numerically by measuring the variance of $D_2$ for a
large sample of pairs of sequences of length $2k - 1$.
Tables of $\Phi(k, t, \eta)$ can be assembled for a range of parameters to
provide a practical way of numerically calculating the accordion contribution.

\subsection{$\Var (D_2)$ for a uniform letter distribution}
\label{section:unif}

For the case of a uniform letter distribution, $\xi_a = 1/d$ for all
$a \in {\cal A}$ where $d = |{\cal A}|$
is the alphabet size, we find that the crabgrass and off-diagonal part of the
accordion contribution to $\Var(D_2)$ are zero, and that an analytic formula
for the remaining, diagonal-accordion, contribution, can easily be found.

\subsubsection{Crabgrass contribution, $\eta = 0$}

When $\eta = 0$, the perturbed binomial distribution reduces to the ordinary
binomial distribution, independent of $c$ \cite{MM04}:
\begin{equation}
g_t(k, 0, c) = {k \choose t} \left(\frac{1}{4}\right)^t \left(\frac{3}{4}\right)^{k - t}.
\end{equation}
Accordingly, the function $f_r({\mathbf W})$ in Eq.~\ref{crabVar} is
independent of the random word $\mathbf W$, its variance is zero, and thus
$\left. \Var (D_2(n_A, n_B, k, t, 0)) \right|_{\rm crabgrass} = 0$.

\subsubsection{Diagonal-accordion contribution}
\label{section:unifacc}

For arbitrary $\eta$ we have (see Fig.~\ref{fig:accordion})
\begin{eqnarray}
\left. \Var (D_2) \right|_{\rm diag. accordion}  & = &
                                   \sum_{u \in I} \sum_{v \in J_u^{ad}} \Cov(Y_u, Y_v)  \nonumber \\
          & = & n_A n_B \sum_{r = -k + 1}^{k - 1} \Cov(Y_u, Y_{u + (r,r)})  \nonumber \\
          & = & n_A n_B \left[ \Cov(Y_u, Y_{u}) +  2\sum_{r = 1}^{k - 1} \Cov(Y_u, Y_{u + (r,r)}) \right].
                                                                                                \nonumber \\
\end{eqnarray}
The covariance is
\begin{equation}
\Cov(Y_u, Y_{u + (r,r)}) = E[Y_u, Y_{u + (r,r)}] - E[Y_u]^2,
\end{equation}
where
\begin{eqnarray}
E[Y_u, Y_{u + (r,r)}] & = & \Prob(Y_u = 1, Y_{u + (r,r)} = 1) \nonumber \\
               & = & \sum_{l = 0}^{\min({k - r,t})} \Prob(\Delta(k - r) = l)
                                                                          \sum_{s_1 = 0}^{t - l} \Prob(\Delta(r) = s_1)
                                                                          \sum_{s_2 = 0}^{t - l} \Prob(\Delta(r) = s_2) \nonumber \\
                & = & \sum_{l = 0}^{\min({k - r,t})} {k - r \choose l} (1 - p_2)^l {p_2}^{k - r - l}
                             \left[ \sum_{s = 0}^{t - l} {r \choose s} (1 - p_2)^s {p_2}^{r - s} \right]^2, \nonumber \\
\label{EYY_diag_acc}
\end{eqnarray}
and
\begin{equation}
E[Y_u] = \Prob(Y_u = 1)
                      = \sum_{l = 0}^t \Prob(\Delta(k) = l)
                      = \sum_{l = 0}^t {k \choose l} (1 - p_2)^l {p_2}^{k - l} .  \label{EY_diag_acc}
\end{equation}
The $l^{\rm th}$ term in Eq.~(\ref{EYY_diag_acc}) accounts for the event that
there are up to $t - l$ mismatches between $(A_i, \ldots, A_{i + r - 1})$ and
$(B_j, \ldots, B_{j + r - 1})$, exactly $l$ mismatches between
$(A_{i + r}, \ldots, A_{i + k - 1})$ and
$(B_{j + r}, \ldots, B_{j + k  - 1})$ and up to $t - l$ mismatches between
$(A_{i + k}, \ldots, A_{i + k + r - 1})$ and
$(B_{j + k}, \ldots, B_{j + k + r - 1})$.

For the case of a uniform letter distribution, one simply sets $p_2 = 1/d$ in
Eqs.~(\ref{EYY_diag_acc}) and (\ref{EY_diag_acc}).

\subsubsection{Off-diagonal-accordion contribution, $\eta = 0$}

The proof that $\left. \Var (D_2) \right|_{\rm off-diag. accordion} = 0$ for a
uniform letter distribution is non-trivial.
First we establish some general results about the distance function
$\Delta(\mathbf{W}_1, \mathbf{W}_2)$, equal to the number of mismatches between
two random $k$-words $\mathbf{W}_1$ and $\mathbf{W}_2$.

For a uniform letter distribution, and for two {\em independent} (i.e.\
non-overlapping) words $\mathbf{W}_1$ and $\mathbf{W}_2$, we have from
Eq.~(\ref{DeltaDistribution})
\begin{equation}
\Prob(\Delta(\mathbf{W}_1, \mathbf{W}_2) = l) = {k \choose l} (1 - p_2)^l {p_2}^{k - l}
         = {k \choose l}\frac{(d - 1)^l}{d^k}.
\end{equation}
If one of the words is fixed to be $\mathbf{w}$, one easily checks that the
conditional probability is also binomial:
\begin{equation}
\Prob(\Delta(\mathbf{W}_1, \mathbf{W}_2) = l | \mathbf{W}_2 = \mathbf{w})  = {k \choose l}\frac{(d - 1)^l}{d^k} = \Prob\left(\Delta(\mathbf{W}_1, \mathbf{W}_2) = l \right).  \label{Prop2a}
\end{equation}
Thus $\Delta(\mathbf{W}_1, \mathbf{W}_2)$ and $\mathbf{W}_2$ are independent
random variables.

Now consider the case of three independent random words
$\mathbf{W}_1$, $\mathbf{W}_2$ and $\mathbf{W}_3$.
Then
\begin{eqnarray}
\lefteqn{\Prob(\Delta(\mathbf{W}_1, \mathbf{W}_2)  =  l_1, \Delta(\mathbf{W}_2, \mathbf{W}_3) = l_2)}  \nonumber \\
& = & \sum_{\mathbf{w} \in {\cal A}^k}
            \Prob(\Delta(\mathbf{W}_1, \mathbf{W}_2) = l_1 | \mathbf{W}_2 = \mathbf{w}) \nonumber \\
&   & \qquad \times
            \Prob(\Delta(\mathbf{W}_2, \mathbf{W}_3) = l_2 | \mathbf{W}_2 = \mathbf{w})
            \Prob(\mathbf{W}_2 = \mathbf{w})   \nonumber \\
& = & \sum_{\mathbf{w} \in {\cal A}^k}
            \Prob(\Delta(\mathbf{W}_1, \mathbf{W}_2) = l_1)
            \Prob(\Delta(\mathbf{W}_2, \mathbf{W}_3) = l_2) \frac{1}{d^k}   \nonumber \\
& = & \Prob(\Delta(\mathbf{W}_1, \mathbf{W}_2) = l_1)
            \Prob(\Delta(\mathbf{W}_2, \mathbf{W}_3) = l_2)  \label{Prop2b}
\end{eqnarray}
where we have used the fact that, once $\mathbf{W}_2$ is fixed,
$\Delta(\mathbf{W}_1, \mathbf{W}_2)$ and $\Delta(\mathbf{W}_2, \mathbf{W}_3)$
depend only on $\mathbf{W}_1$ and $\mathbf{W}_3$ respectively, and so are
effectively independent.

We now generalise Eqs.~(\ref{Prop2a}) and (\ref{Prop2b}) to the following
proposition $P_N$, which will be proved by induction: \\
For given $N \ge 2$, let $\mathbf{W}_1, \ldots, \mathbf{W}_{N + 1}$ be mutually
independent $k$-words, and define
\begin{equation}
\Delta_i = \Delta(\mathbf{W}_i, \mathbf{W}_j), \qquad i = 1,\ldots, N.
\end{equation}
Then for any $\mathbf{w} \in {\cal A}^k$,
\begin{eqnarray}
\lefteqn{\Prob\left(\Delta_1 = l_1, \ldots, \Delta_{N - 1} = l_{N - 1} | \mathbf{W}_N = \mathbf{w} \right)}
                                       \nonumber \\
& = & \Prob\left(\Delta_1 = l_1, \ldots, \Delta_{N - 1} = l_{N - 1}\right)  \label{PropNa}
\end{eqnarray}
and
\begin{eqnarray}
\lefteqn{\Prob\left(\Delta_1 = l_1, \ldots, \Delta_{N} = l_{N} \right)}
                                       \nonumber \\
& = & \Prob\left(\Delta_1 = l_1, \ldots, \Delta_{N - 1} = l_{N - 1}\right)  \Prob\left(\Delta_{N} = l_{N} \right).
                                        \label{PropNb}
\end{eqnarray}

Note that Eq.~(\ref{PropNa}) could equivalently be written as
\begin{eqnarray}
\lefteqn{\Prob\left(\Delta_1 = l_1, \ldots, \Delta_{N - 1} = l_{N - 1} | \mathbf{W}_N \in R \right)}
                                       \nonumber \\
& = & \Prob\left(\Delta_1 = l_1, \ldots, \Delta_{N - 1} = l_{N - 1}\right)  \label{PropNaAlt},
\end{eqnarray}
where $R \subset {\cal A}^k$ is any restricted set of $k$-words.
Note also that combining Eq.~(\ref{PropNb}) for the propositions $P_2$ to $P_N$
implies
\begin{equation}
\Prob\left(\Delta_1 = l_1, \ldots, \Delta_{N} = l_{N} \right) =
 \Prob\left(\Delta_1 = l_1\right) \times \ldots \times \Prob\left(\Delta_{N} = l_{N} \right)
                                        \label{PropNbAlt}
\end{equation}

The proposition $P_2$ is proved by Eqs.(\ref{Prop2a}) and (\ref{Prop2b}).
It remains to prove that $P_N$ implies $P_{N + 1}$.
Define $S(\mathbf{w},l) = \{\mathbf{x} \in {\cal A}^k | \Delta(\mathbf{x},\mathbf{w}) = l\}$.
Starting with the left hand side of Eq.~(\ref{PropNa}) with
$N$ replaced by $N + 1$, we have
\begin{eqnarray}
\lefteqn{\Prob\left(\Delta_1 = l_1, \ldots, \Delta_{N } = l_{N} | \mathbf{W}_{N + 1} = \mathbf{w} \right)}
                                       \nonumber \\
& = & \Prob\left(\Delta_1  =  l_1, \ldots, \Delta_{N } = l_{N} | \mathbf{W}_{N + 1} = \mathbf{w}, \mathbf{W}_N \in S(\mathbf{w}, l_N) \right) \nonumber \\
& & \qquad \times \Prob\left(\mathbf{W}_N \in S(\mathbf{w}, l_N) \right)
                                       \nonumber \\
& & + \Prob\left(\Delta_1  =  l_1, \ldots, \Delta_{N } = l_{N} | \mathbf{W}_{N + 1} = \mathbf{w}, \mathbf{W}_N \notin S(\mathbf{w}, l_N) \right) \nonumber \\
& & \qquad \times \Prob\left(\mathbf{W}_N \notin S(\mathbf{w}, l_N) \right)
                                       \nonumber \\
& = & \Prob\left(\Delta_1  =  l_1, \ldots, \Delta_{N - 1} = l_{N - 1} | \mathbf{W}_{N + 1} = \mathbf{w}, \mathbf{W}_N \in S(\mathbf{w}, l_N) \right) \nonumber \\
& & \qquad \times \Prob\left(\mathbf{W}_N \in S(\mathbf{w}, l_N) \right),
                                       \nonumber
\end{eqnarray}
where the second term is zero since
``$\Delta_{N } = l_{N}$'' and ``$\mathbf{W}_N \notin S(\mathbf{w}, l_N)$''
are mutually exclusive events, and the requirement ``$\Delta_{N } = l_{N}$''
has been dropped from the first term since it is automatically satisfied by the
condition
``$\mathbf{W}_{N + 1} = \mathbf{w}$ and $\mathbf{W}_N \in S(\mathbf{w}, l_N)$''.
Then, since $\Delta_1, \ldots, \Delta_{N - 1}$ are independent of
$\mathbf{W}_{N + 1}$,  and rewriting the second factor, we have
\begin{eqnarray}
\lefteqn{\Prob\left(\Delta_1 = l_1, \ldots, \Delta_{N } = l_{N} | \mathbf{W}_{N + 1} = \mathbf{w} \right)}
                                       \nonumber \\
& = & \Prob\left(\Delta_1  =  l_1, \ldots, \Delta_{N - 1} = l_{N - 1} | \mathbf{W}_N \in S(\mathbf{w}, l_N) \right) \nonumber \\
& & \qquad \times \Prob\left(\Delta_N = l_N | \mathbf{W}_{N + 1} = \mathbf{w}\right) \nonumber \\
& = & \Prob\left(\Delta_1  =  l_1, \ldots, \Delta_{N - 1} = l_{N - 1}\right)  \Prob\left(\Delta_N = l_N\right)
\qquad \mbox{by Eqs.(\ref{Prop2a}) and (\ref{PropNaAlt})} \nonumber \\
& = & \Prob\left(\Delta_1  =  l_1, \ldots, \Delta_{N} = l_{N }\right) \qquad \mbox{by Eq.~(\ref{PropNb})}
\label{PropNplus1a}
\end{eqnarray}
which establishes the first part of proposition $P_{N+1}$.
Starting with the left hand side of Eq.~(\ref{PropNb}) with
$N$ replaced $N + 1$,
\begin{eqnarray}
\lefteqn{\Prob\left(\Delta_1 = l_1, \ldots, \Delta_{N + 1} = l_{N + 1} \right)}
                                       \nonumber \\
& = & \sum_{\mathbf{w} \in {\cal A}^k}
\Prob\left(\Delta_1 = l_1, \ldots, \Delta_{N} = l_{N} | \mathbf{W}_{N + 1} = \mathbf{w}\right)
                                       \nonumber \\
& & \qquad \times \Prob\left(\Delta_{N + 1} = l_{N + 1} | \mathbf{W}_{N + 1} = \mathbf{w}\right)
                              \Prob\left(\mathbf{W}_{N + 1} = \mathbf{w}\right)
                                       \nonumber \\
& = & \sum_{\mathbf{w} \in {\cal A}^k}
\Prob\left(\Delta_1 = l_1, \ldots, \Delta_{N} = l_{N}\right)
 \Prob\left(\Delta_{N + 1} = l_{N + 1}\right)
                              \frac{1}{d^k} \qquad \mbox{by Eq.~(\ref{PropNplus1a})} \nonumber \\
& = & \Prob\left(\Delta_1 = l_1, \ldots, \Delta_{N} = l_{N}\right)
 \Prob\left(\Delta_{N + 1} = l_{N + 1}\right),
\end{eqnarray}
which establishes the second half of proposition $P_{N + 1}$.
\footnote{Aside: For an alternate proof that
$\left. \Var (D_2(n_A, n_B, k, t, 0)) \right|_{\rm crabgrass} = 0$
one can apply the above proposition to the third line of Eq.~(5) of
\cite{BKW08}.}

\begin{figure}
   \centering
   \includegraphics[height=8.1cm]{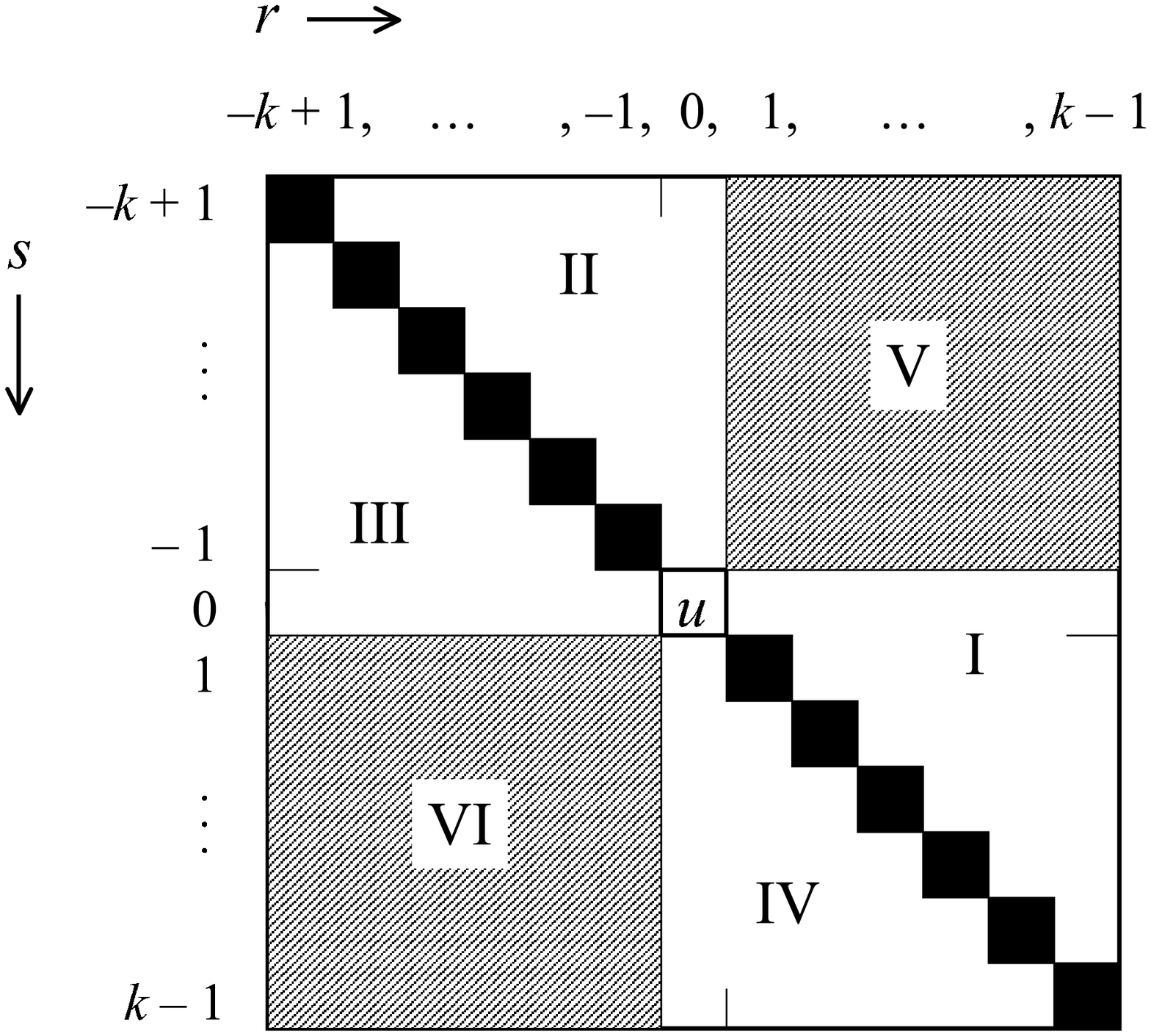}
   \caption{The main diagonal and off-diagonal regions I to VI of $J_u^a$.}
   \label{fig:accordion}
\end{figure}

We are now in a position to calculate
\begin{equation}
\left. \Var (D_2) \right|_{\rm off-diag. accordion}  =
                                   \sum_{u \in I} \sum_{v \in J_u^{ao}} \Cov(Y_u, Y_v).
\end{equation}
Writing $u = (i, j)$, $v = (i + r, j + s) \in J_u^{ao}$, the off-diagonal part
$J_u^{ao}$ can be subdivided into six parts illustrated in
Fig.~\ref{fig:accordion}, namely
\begin{itemize}
\item[I:] {$0 \le s < r \le k - 1$};
\item[II:] {$-k + 1 \le s < r \le 0$};
\item[III:] {$-k + 1 \le r < s \le 0$};
\item[IV:] {$0 \le r < s \le k - 1$};
\item[V:] {$1 \le r \le k - 1$, $-k +1 \le s \le -1$};
\item[VI:] {$1 \le s \le k - 1$, $-k +1 \le r \le -1$}.
\end{itemize}
We proceed to prove that $\Cov(Y_u, Y_v)$ vanishes for each of the six cases.

\begin{figure}
  \centering
   \includegraphics[height=11.5cm]{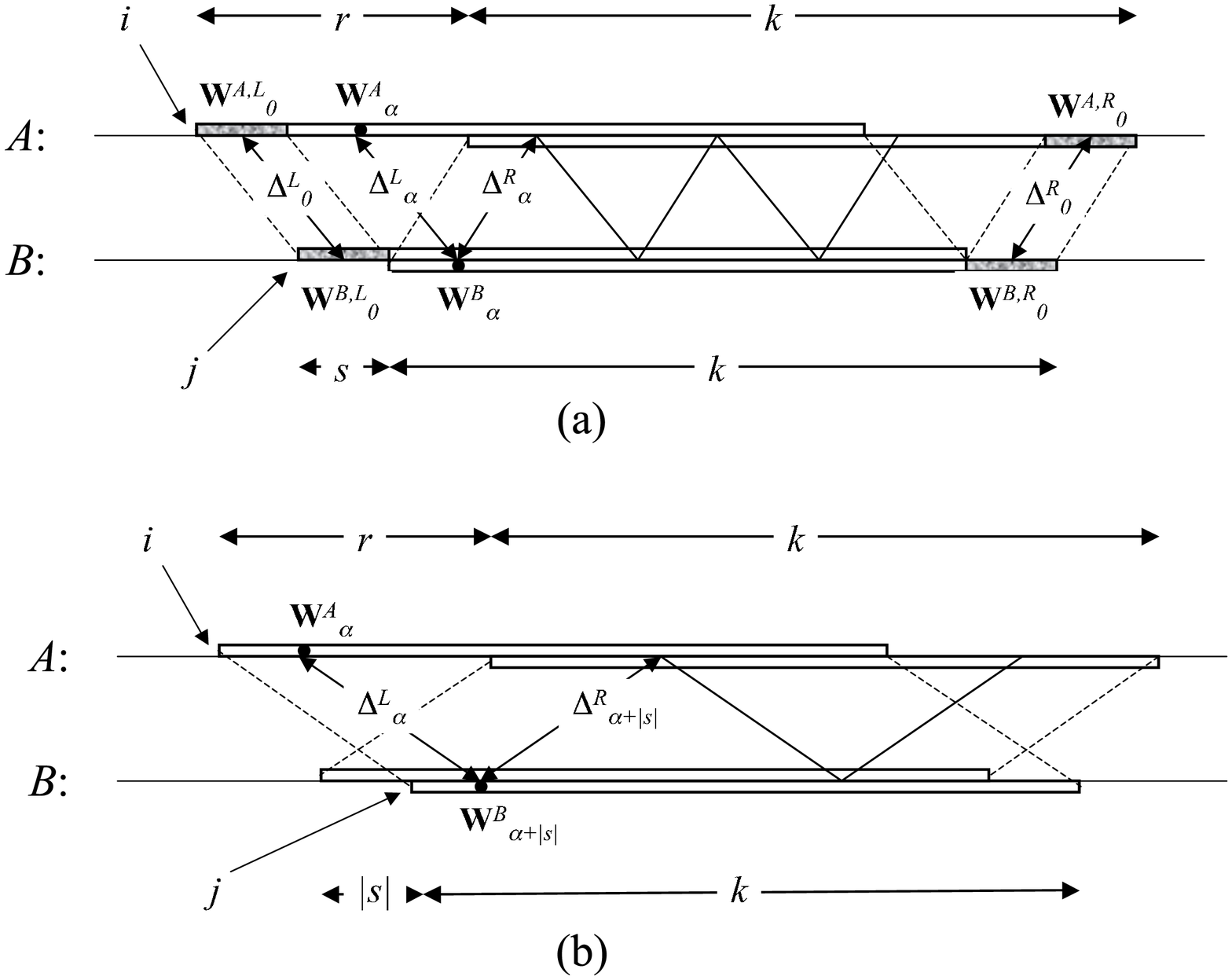}
   \caption{(a) Case I of the off-diagonal accordion contribution to $\Var(D_2)$.
   Cases II, III and IV are obtained by reflection or by interchanging the roles of $A$ and $B$.
   (b) Case V of the off-diagonal contribution.
   Case VI is obtained by interchanging the roles of $A$ and $B$.}
   \label{fig:offDiag}
\end{figure}

Case I is illustrated in Fig.~\ref{fig:offDiag}(a).
The union of the overlapping words
$\mathbf{W}_u^A = (A_i,\ldots, A_{i + k - 1})$ and
$\mathbf{W}_v^A = (A_{i + r},\ldots, A_{i + k + r - 1})$
is subdivided into the shaded pieces
$\mathbf{W}_0^{A,L} = (A_i, \ldots, A_{i + s - 1})$ and
$\mathbf{W}_0^{A,R} = (A_{i + k + r - s}, \ldots, A_{i + k + r - 1})$
each of length $s$, and a set of and a set of single-letter words
$\mathbf{W}_\alpha^{A} = (A_{i + s + \alpha -1})$, $\alpha = 1,\ldots, k + r - 2s$.

Similarly, the union of the overlapping words
$\mathbf{W}_u^B = (B_j,\ldots, B_{j + k - 1})$ and
$\mathbf{W}_v^B = (B_{j + s},\ldots, B_{j + k + s - 1})$
is subdivided into the shaded pieces
$\mathbf{W}_0^{B,L} = (B_i, \ldots, B_{j + s - 1})$ and
$\mathbf{W}_0^{B,R} = (B_{j + k}, \ldots, B_{j + k + s - 1})$
each of length $s$, and a set of and a set of single-letter words
$\mathbf{W}_\alpha^{B} = (B_{j + s + \alpha -1})$, $\alpha = 1,\ldots, k - s$.

Define
\begin{equation}
\Delta^L_0 = \Delta(\mathbf{W}_0^{A,L}, \mathbf{W}_0^{B,L}),  \qquad
\Delta^R_0 = \Delta(\mathbf{W}_0^{B,R}, \mathbf{W}_0^{A,R}) \nonumber
\end{equation}
\begin{equation}
\Delta^L_\alpha = \Delta(\mathbf{W}_\alpha^A, \mathbf{W}_\alpha^B),  \qquad
\Delta^R_\alpha = \Delta(\mathbf{W}_\alpha^B, \mathbf{W}_{\alpha + r - s}^A),
\qquad \alpha = 1,\ldots, k - s.  \nonumber
\end{equation}
Then
\begin{equation}
\Delta(\mathbf{W}_u^A, \mathbf{W}_u^B) = \sum_{\alpha = 0}^{k  - s} \Delta_\alpha^L, \qquad
\Delta(\mathbf{W}_v^A, \mathbf{W}_v^B) = \sum_{\alpha = 0}^{k  - s} \Delta_\alpha^R.
\end{equation}

With the indicator variables $Y_u$ and $Y_v$ defined as above, we have
\begin{eqnarray}
\lefteqn{E(Y_u, Y_v) = \Prob(Y_u = 1, Y_v = 1)} \nonumber \\
& = & \Prob\left(\Delta(\mathbf{W}_u^A, \mathbf{W}_u^B) \le t,
                            \Delta(\mathbf{W}_v^A, \mathbf{W}_v^B) \le t \right)   \nonumber \\
& = & \sum_{\{m_0, \ldots, m_{k - s}\}\in I_t} \sum_{\{l_0, \ldots, l_{k - s}\}\in I_t}
                         \Prob\left(\Delta_0^L = m_0, \ldots \right. \nonumber \\
& & \qquad  \left.    \ldots, \Delta_{k - s}^L = m_{k - s},
                                          \Delta_0^R = l_0, \ldots, \Delta^R_{k - s} = l_{k - s} \right), \label{EYuYvCaseI}
\end{eqnarray}
where the index set summed over is
\begin{equation}
I_t = \left\{l_0, \ldots, l_{k - s} \left|\, 0 \le l_0 \le s, 0 \le l_1, \ldots, l_{k - s} \le 1,
           \sum_{\alpha = 0}^{k - s} l_\alpha \le t \right. \right\}.
\end{equation}
The set
$\{ \Delta_1^L, \ldots, \Delta_{k - s}^L, \Delta_1^R, \ldots, \Delta_{k - s}^R \}$
partitions into a collection of disjoint subsets of the form
$\{\Delta_\alpha^L, \Delta_\alpha^R, \Delta_{\alpha + r - s}^L, \Delta_{\alpha + r - s}^R, \Delta_{\alpha + 2(r - s)}^L, \ldots \}$,
$\alpha = 1, \ldots, r - s$ (indicated by the zig-zag line in
Fig.~\ref{fig:offDiag}(a)),
each of which satisfies the conditions of the proposition $P_N$ for some $N$.
Also, these subsets are independent of one another and of $\Delta_0^L$ and
$\Delta_0^R$, since they contain random variables which are functions of
corresponding disjoint subsets of letters.

Thus we can factor the probability in Eq.(\ref{EYuYvCaseI}) and rearrange the
sum to obtain
\begin{eqnarray}
E(Y_u, Y_v) & = &  \sum_{\{m_0, \ldots, m_{k - s}\}\in I_t} \Prob\left(\Delta_0^L = m_0\right)
                                                       \ldots \Prob\left(\Delta_{k - s}^L = m_{k - s}\right)  \nonumber \\
& & \qquad \times \sum_{\{l_0, \ldots, l_{k - s}\}\in I_t} \Prob\left(\Delta_0^R = l_0\right)
                                                       \ldots \Prob\left(\Delta_{k - s}^R = l_{k - s}\right)   \nonumber \\
& = &  \sum_{\{m_0, \ldots, m_{k - s}\}\in I_t} \Prob\left(\Delta_0^L = m_0,
                                                       \ldots, \Delta_{k - s}^L = m_{k - s}\right)  \nonumber \\
& & \qquad \times \sum_{\{l_0, \ldots, l_{k - s}\}\in I_t} \Prob\left(\Delta_0^R = l_0,
                                                       \ldots, \Delta_{k - s}^R = l_{k - s}\right)   \nonumber \\
& = & \Prob\left(\Delta(\mathbf{W}_u^A, \mathbf{W}_u^B) \le t \right)
           \Prob\left(\Delta(\mathbf{W}_v^A, \mathbf{W}_v^B) \le t \right)   \nonumber \\
& = & E(Y_u) E(Y_v).
\end{eqnarray}
Thus $\Cov(Y_u, Y_v) = 0$ for $v$ in the Case I part of $J_u^{ao}$.
Cases II, III and IV can be similarly dealt with by reversing the order of both
sequences, interchanging the roles of sequences $A$ and $B$, or both.

Case V is illustrated in Fig.~\ref{fig:offDiag}(b).
This time the union of the overlapping words
$\mathbf{W}_u^A$ and $\mathbf{W}_v^A$
is subdivided into the set of single-letter words
$\mathbf{W}_\alpha^{A} = (A_{i + \alpha - 1})$,
$\alpha = 1,\ldots, k + r$, and the union of the overlapping words
$\mathbf{W}_u^B$ and $\mathbf{W}_v^B$ is subdivided into the set of
single-letter words
$\mathbf{W}_\alpha^{B} = (B_{j -|s| + \alpha -1})$, $\alpha = 1,\ldots, k + |s|$.
We define
\begin{equation}
\Delta^L_\alpha = \Delta(\mathbf{W}_\alpha^A, \mathbf{W}_{\alpha + |s|}^B),  \qquad
\Delta^R_\alpha = \Delta(\mathbf{W}_\alpha^B, \mathbf{W}_{\alpha + r}^A),
\qquad \alpha = 1,\ldots, k.  \nonumber
\end{equation}
Then
\begin{equation}
\Delta(\mathbf{W}_u^A, \mathbf{W}_u^B) = \sum_{\alpha = 1}^k \Delta_\alpha^L, \qquad
\Delta(\mathbf{W}_v^A, \mathbf{W}_v^B) = \sum_{\alpha = 1}^k \Delta_\alpha^R,
\end{equation}
and
\begin{eqnarray}
E(Y_u, Y_v) & = & \Prob(Y_u = 1, Y_v = 1) \nonumber \\
& = & \Prob\left(\Delta(\mathbf{W}_u^A, \mathbf{W}_u^B) \le t,
                            \Delta(\mathbf{W}_v^A, \mathbf{W}_v^B) \le t \right)   \nonumber \\
& = & \sum_{\{m_1, \ldots, m_k\}\in I_t} \sum_{\{l_1, \ldots, l_k\}\in I_t}
                         \Prob\left(\Delta_1^L = m_1, \ldots \right. \nonumber \\
& & \qquad  \left.    \ldots, \Delta_k^L = m_k,
                                          \Delta_1^R = l_1, \ldots, \Delta^R_k = l_k \right), \label{EYuYvCaseV}
\end{eqnarray}
where the index set is now
\begin{equation}
I_t = \left\{l_0, \ldots, l_k \left|\, 0 \le l_1, \ldots, l_k \le 1,
           \sum_{\alpha = 1}^k l_\alpha \le t \right. \right\}.
\end{equation}

The set
$\{ \Delta_1^L, \ldots, \Delta_k^L, \Delta_1^R, \ldots, \Delta_k^R \}$
partitions into a collection of disjoint subsets of the form
$\{\Delta_\alpha^L, \Delta_{\alpha + |s|}^R, \Delta_{\alpha + r + |s|}^L, \ldots \}$,
$\alpha = 1, \ldots, r$, or\\
$\{\Delta_\alpha^R, \Delta_{\alpha + |s|}^L, \Delta_{\alpha + r + |s|}^R, \ldots \}$,
$\alpha = 1, \ldots, |s|$
(indicated by the zig-zag line in Fig.~\ref{fig:offDiag}(b)),
each of which satisfies the conditions of the proposition $P_N$ for some $N$,
and which are mutually independent.
Thus we can factor the probability in Eq.(\ref{EYuYvCaseV}), rearrange the sum
and recombine the probabilities to obtain
\begin{eqnarray}
E(Y_u, Y_v) & = &  \sum_{\{m_1, \ldots, m_k\}\in I_t} \Prob\left(\Delta_1^L = m_1,
                                              \ldots, \Delta_k^L = m_k\right)  \nonumber \\
& & \qquad \times \sum_{\{l_1, \ldots, l_k\}\in I_t} \Prob\left(\Delta_1^R = l_1,
                                              \ldots, \Delta_k^R = l_k\right)   \nonumber \\
& = & \Prob\left(\Delta(\mathbf{W}_u^A, \mathbf{W}_u^B) \le t \right)
           \Prob\left(\Delta(\mathbf{W}_v^A, \mathbf{W}_v^B) \le t \right)   \nonumber \\
& = & E(Y_u) E(Y_v),
\end{eqnarray}
giving  $\Cov(Y_u, Y_v) = 0$ for $v$ in the Case V part of $J_u^{ao}$.
Case VI can be similarly dealt with by interchanging the roles of sequences $A$
and $B$.

\section*{Acknowledgement}

We thank Joerg Arndt for help with optimising the code of the simulations.  We
also would like to express our thanks to anonymous referees for valuable
suggestions which have improved the content of this paper.  This work was
funded in part by ARC discovery grant DP0559260.

%

\end{document}